\documentclass[twocolumn,letter]{revtex4}
\usepackage{amsmath}
\usepackage{graphicx}
\usepackage{verbatim}
\usepackage{subfigure}
\usepackage{setspace}

\begin{document}
\title{On-demand single-electron transfer between distant quantum dots}
\author{R.~P.~G.~McNeil}
\author{M.~Kataoka\footnote{Current address: National Physical Laboratory, Hampton Rd, Teddington, Middlesex, TW11 0LW,
United Kingdom.}}
\author{C.~J.~B.~Ford}
\author{C.~H.~W.~Barnes}
\author{D.~Anderson}
\author{G.~A.~C.~Jones}
\author{I.~Farrer}
\author{D.~A.~Ritchie}
\affiliation{Cavendish Laboratory, University of Cambridge, JJ Thomson Avenue, Cambridge CB3 0HE, United Kingdom }

\maketitle

 Single-electron circuits of the future, consisting of a network of quantum dots, will require a mechanism to transport electrons from one functional part to another.  For example, in a quantum computer\cite{Loss1998} decoherence and circuit complexity can be reduced by separating qubit manipulation from measurement and by providing some means to transport electrons from one to the other.\cite{Barnes2000}
 Tunnelling between neighbouring dots has been demonstrated\cite{Petta2005,Pioro-Ladriere2008} with great control, and the manipulation of electrons in single and double-dot systems is advancing rapidly.\cite{Elzerman2004, Hanson2005,Morello2010,Hanson2007}
For distances greater than a few hundred nanometres neither free propagation nor tunnelling are viable whilst maintaining confinement of single electrons.
 Here we show how a single electron may be captured in a surface acoustic wave minimum and transferred from one quantum dot to a second unoccupied dot along a long empty channel.
 The transfer direction may be reversed and the same electron moved back and forth over sixty times without error, a cumulative distance of $0.25$\,mm.
 Such on-chip transfer extends communication between quantum dots to a range that may allow the integration of discrete quantum information-processing components and devices.

\begin{figure*}[tbp]
\begin{center}
\includegraphics[width=17.5cm]{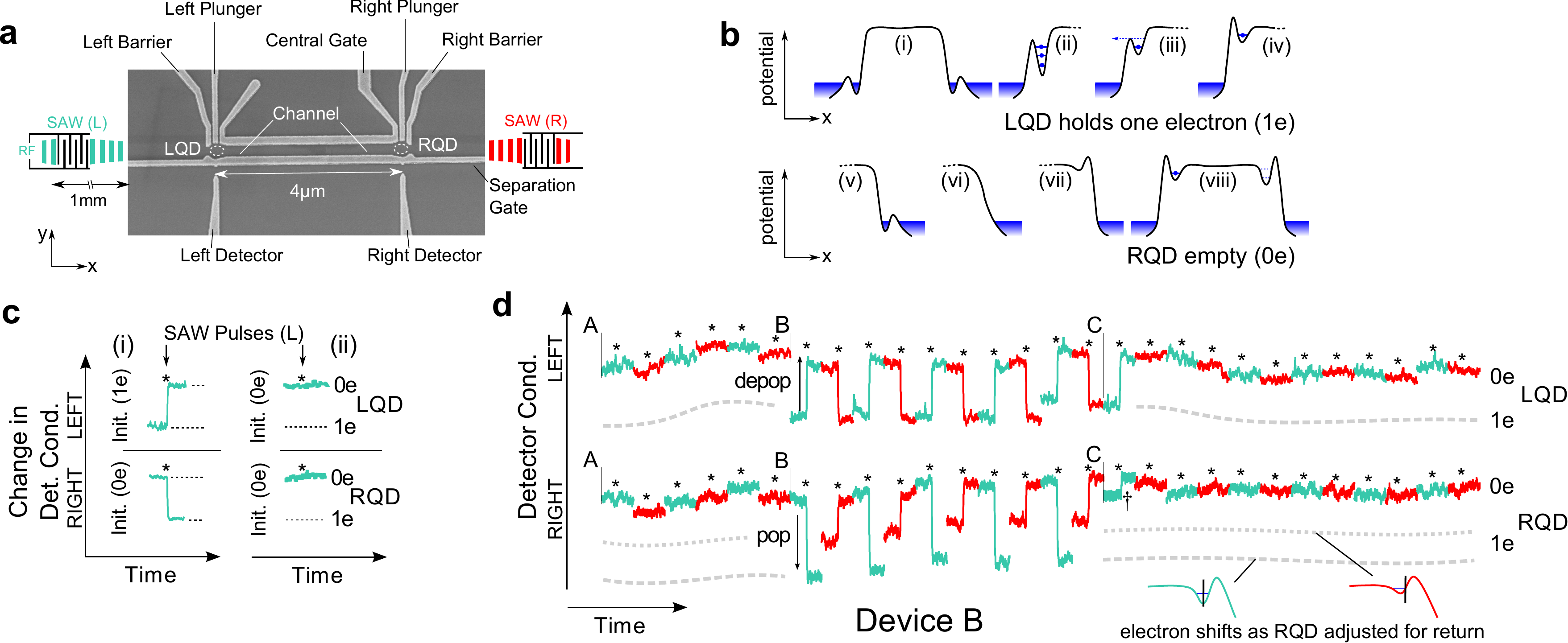}
\caption{\label{FIG1} \textbf{Device, initialisation and single-electron transfer.} \textbf{a}, Scanning electron micrograph of device. Voltages applied to gates (light grey) create QDs (dashed circles) connected by $4$\,$\mu$m channel. Applying microwave (RF) pulse to left/right transducer, (placed $1$\,mm from device),  generates SAW pulses which trap and transport electrons.
\textbf{b}, Schematic of potential between left and right QDs during initialisation of LQD with one electron ($1$e) (i--iv) and then RQD with no electrons ($0$e) (v--viii).
\textbf{c}, Change in detector conductance when SAW pulse ($\ast$) is applied to system set up as in \textbf{b}(viii). Empty RQD is populated when electron leaves LQD. Second pair of traces shows control case with LQD starting empty (0e)(traces are 1\,s long).
\textbf{d}, Single-electron rally: Dots and channel initialised empty before A. (A-B) series of control pulses to verify that system is empty; (B-C) two-way transfer of single electron between QDs; (C) electron removed from system with clearing pulse. SAW pulse duration $300$\,ns. Small step marked ($\dagger$) is random switching event and not SAW driven. Time between traces is not plotted.
}
\end{center}
\end{figure*}

 Our device consists of two quantum dots (QDs) connected by a long channel, as shown in Fig.~\ref{FIG1}a.
 Negative voltages applied to patterned metal surface gates deplete a two-dimensional electron gas (2DEG) that lies $90$\,nm below the surface. The voltages are chosen so that the potential of the system is above the Fermi energy, and in thermal equilibrium the dots and channel contain no electrons.

The QDs are adjusted by the two plunger and barrier gates. The plunger raises and lowers the dot while the barrier controls the degree of isolation between the dot and the neighbouring reservoir. The charge in each QD is detected by its effect on the conductance of high resistance constrictions\cite{Field1993} on the other side of a narrow ``separation'' gate.
A single electron can be initialised in one QD (Fig.~\ref{FIG1}b(iv)), and then transferred at will to the other QD using a short burst of surface-acoustic-waves (SAWs). In a piezoelectric material (like GaAs) SAWs create a moving potential modulation which can trap and transport electrons. The transferred electron can be returned using a second SAW pulse travelling from the opposite direction giving two-way transfer.

Initialisation of the dots is shown schematically in Fig.~\ref{FIG1}b. To set up an occupied left dot (LQD), the left barrier (LB) and plunger (LP) gates are lowered to fill LQD (i), LB is then raised to isolate LQD from the reservoir (ii) and LP raised to selectively depopulate the dot leaving one electron (iii), or more if desired (see Fig.~S1). LB and LP can then be stepped to their final voltages (iv). The dot now contains a chosen number of electrons held close to, but below, the channel potential. An empty dot is initialised in a similar way but with the plunger being raised first (Fig.~\ref{FIG1}b(v-viii)). The final voltages for both the empty and occupied QDs (iv/viii) are the same and thus detector conductance indicates the number of electrons in each dot (see supplementary information).

On-demand depopulation of an initialised QD is achieved by a brief SAW pulse.
Applying a microwave signal to the left transducer generates a SAW. The accompanying potential modulation, moving at $2870$\, ms$^{-1}$, captures the electron from LQD and transfers it in $1.4$\,ns to RQD.
Figure~\ref{FIG1}c(i) shows the conductance of the left and right detectors for an occupied LQD (1e) and an unoccupied RQD (0e) when a  SAW pulse ($300$\,ns long) is sent from the left (SAW(L)). The transfer of charge is shown by the simultaneous step change in conductance of the detectors.

During the SAW pulse sequences the QDs are not simply exchanging electrons with their neighbouring reservoirs (counter to the SAW propagation direction) since for the control case, with an empty starting dot, no change in the detector conductance is seen (Fig.~\ref{FIG1}c(ii)).
Another possibility is a ``Newton's cradle'' arrangement where an electron from one dot moves into the channel, causing a series of electrons in traps along the channel to shuffle up, ejecting the last electron into the second dot.
However, the SAW amplitude is $2.5$ times greater than that at which electrons are caught in the channel and so there are no electrons to be ``shuffled along''.
Thus in Fig.~\ref{FIG1}c(i) a single electron is being transferred between the dots.

The two transducers allow for bidirectional transfer between the QDs and single electrons (or pairs) can be ``batted'' backwards and forwards in a game of ``ping-pong'', with rallies of tens to hundreds of cycles. Figure.~\ref{FIG1}d is an example of such a single-electron rally. Both QDs are emptied before A, and six control pulses (three SAW(L)-SAW(R) pairs) show the system to be empty. At B an electron is loaded into LQD. The electron is then sent back and forth by ten alternate SAW pulses (five pairs) until at C the RQD barrier is partially lowered and a ``clearing'' pulse removes the electron from the channel -- in this case to the right reservoir but potentially into the next section of a QD circuit. The small step in the right detector signal ($\dagger$) is a random switching event near the detector. It is not coincident with the SAW pulse but occurs $50$\,ms later. No further electron movement is seen in the subsequent ten pulses.

\begin{figure}[t]
\begin{center}
 \includegraphics[width=8.5cm]{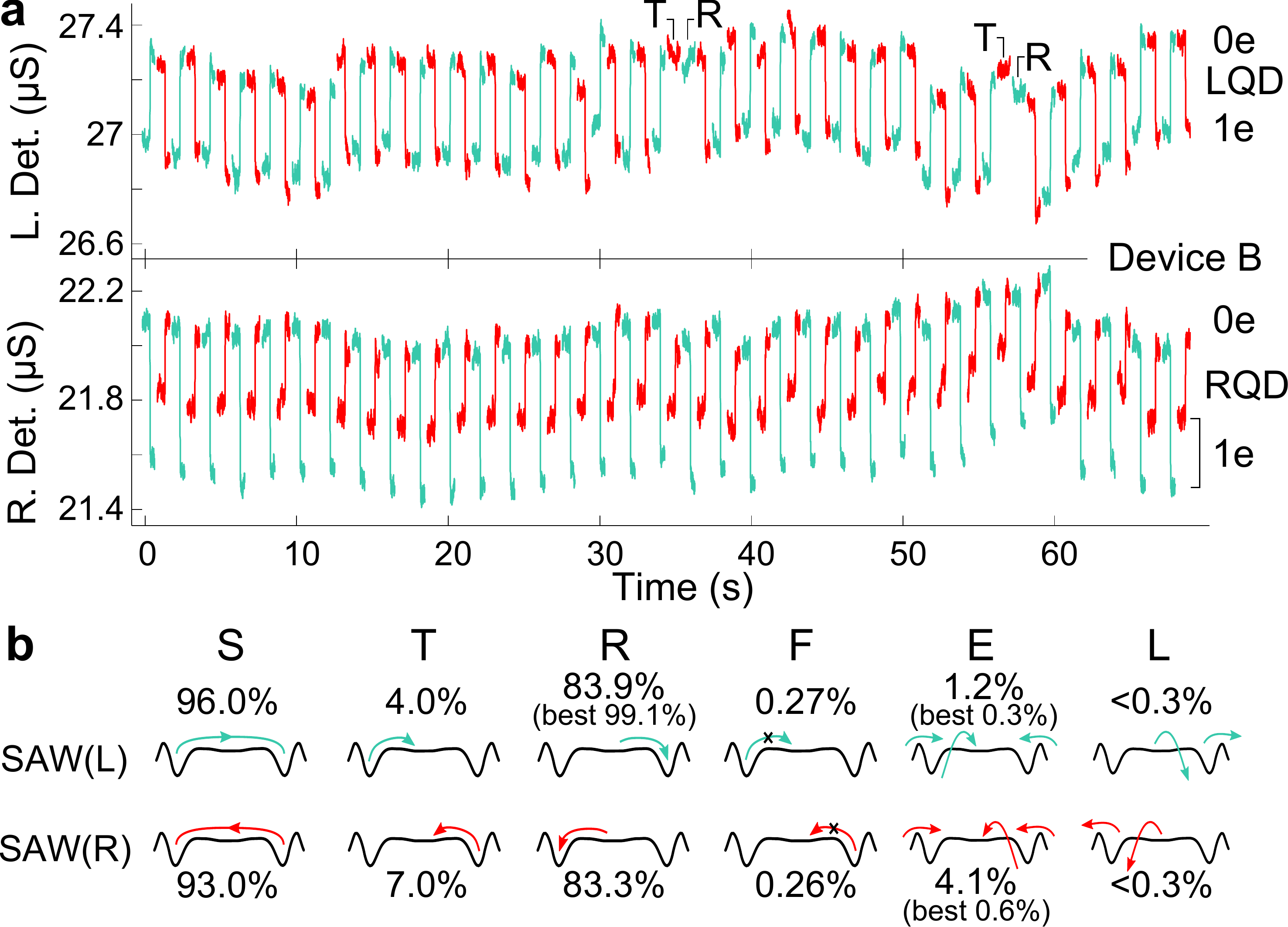}\\
  \caption{\textbf{Single-electron transfer reliability.}
\textbf{a},  Example of bidirectional electron transfer.  Electron is transferred between QDs $35$ times before getting trapped in channel (T). Next SAW(L) pulse recovers electron (R). Time between traces not plotted, SAW pulse duration $300$\,ns.
\textbf{b}, Transfer statistics for full data set (excerpt seen in \textbf{a}). Probability of events shown: ideal transfer (S), depopulation to channel trap (T), recovery from channel (R), failure to depopulate (F), arrival of additional electron (E), loss of electron from system (L) for left and right pulses. (Values in brackets are for different voltages.)}\label{FIG2}\label{Fig:Stats}
\end{center}
\end{figure}

In this device rallies of over $60$ pulses were possible with a single electron going back and forth between the QDs. A run of $35$ transfers is shown in Fig.~\ref{FIG2}a with the statistics of the wider data set in Fig.~\ref{FIG2}b.
Rallies are broken when the transfer fails, which occurred in one of two ways. Occasionally depopulation of the starting dot fails (F) (see Fig.~\ref{FIG2}b), in which case no electron arrives in the second dot. The chances of this can be reduced by raising the starting dot, towards the channel potential, or by increasing the SAW amplitude, although larger SAWs can pose problems, e.g. lifting the transferred electron over the barrier of the second dot. Given successful depopulation, the transfer may still fail if the electron becomes trapped in the channel (T). This type of failure was more common with pulses from the weaker right transducer and examples  can be seen in Fig.~\ref{FIG2}a, (marked T). Here a SAW(R) pulse fails to transfer the electron all the way to LQD, leaving it trapped in the channel. However, the next pulse from the other transducer recovers this electron, returning it to RQD. The lower probabilities of recovery (R) compared with transfer (S) indicate that electrons trapped in the channel may relax deeper into impurity traps than electrons that are carried through in SAW minima. This second type of error can also occur in another way (X), described later, though this can be eliminated by lowering the potential in the second dot.

A third error mechanism (E) is the arrival of an additional electron, which is then transferred with the initial electron. Electrons are seen to enter the system during pulses from the right transducer which may have been caused or exacerbated by adjusting RQD before the SAW(R) pulses. No electrons appeared in the system during SAW(L) pulses. Increasing the isolation of the QDs and the channel from the surrounding reservoirs will reduce this. In none of the traces is the electron seen to leave the system (L).

\begin{figure}[tb]
\begin{center}
\includegraphics[width = 8.5cm]{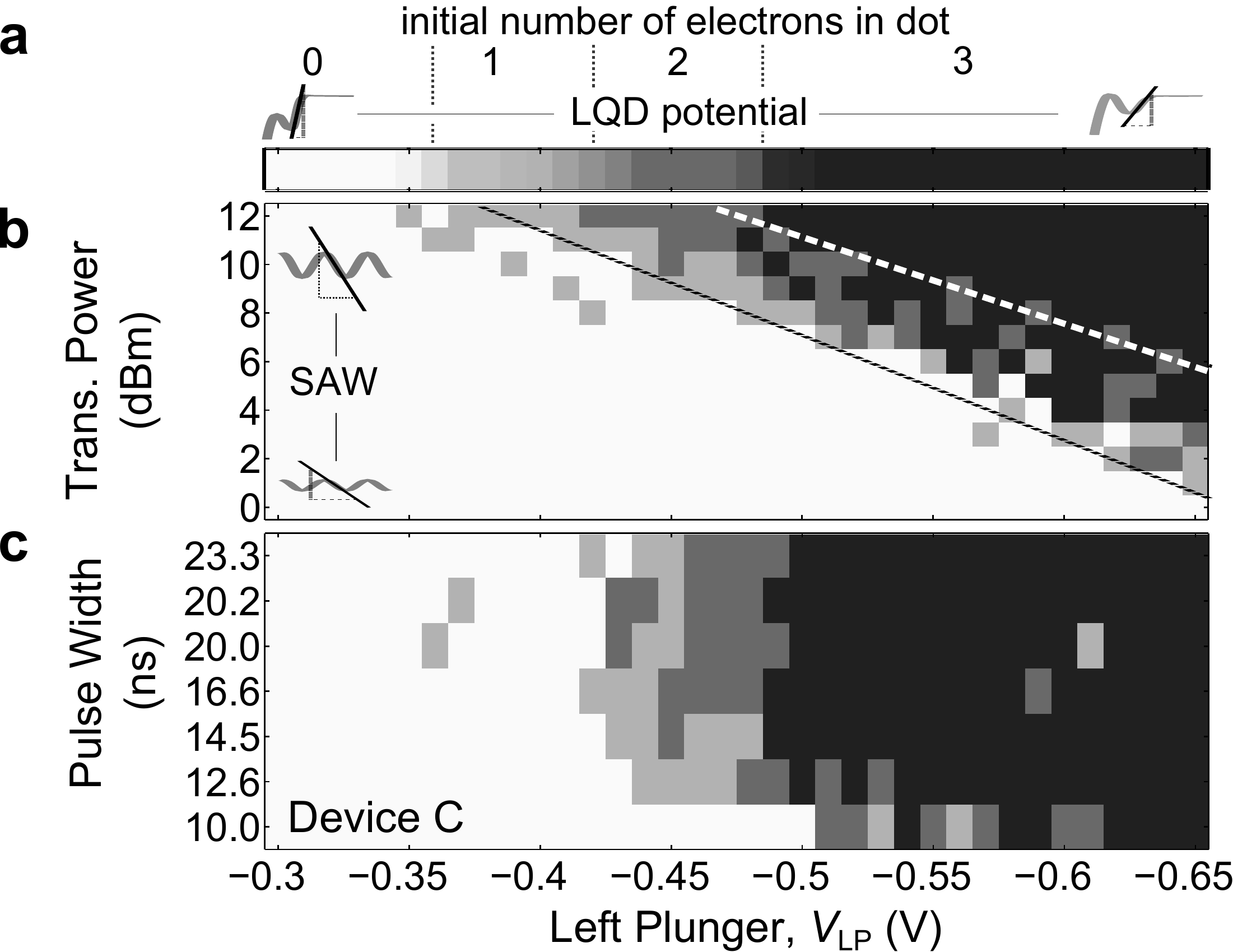}
\caption{\label{FIG5}\textbf{SAW-power and pulse-width dependence of LQD depopulation.}
\textbf{a}, Numbers: initial population of LQD \textit{vs} plunger voltage $V_{\textrm{LP}}$. Greyscale also used as key for shading in \textbf{b} \& \textbf{c} to indicate number of electrons removed by pulse.
\textbf{b}, Depopulation of LQD at different SAW(L) transducer powers, pulse width $100$\,ns. The relative slope of SAW and potential determine if depopulation occurs.
\textbf{c}, Depopulation of LQD at different pulse widths for SAW(L) power of $11$\,dBm. Pulse widths shorter than 14.5\,ns do not achieve full amplitude and so depopulation fails at smaller $V_{LP}$. By comparison with \textbf{b} the peak SAW power at $10$ and $12.6$\,ns can be estimated as $7$ and $10$\,dBm respectively. Pulse widths are measured values of RF source and not linearly spaced. }
\end{center}
\end{figure}

The ability of the SAW to transport electrons depends on the SAW amplitude relative to the potential.\cite{Kataoka2006PRB,Rahman2006} Removing an electron from the starting dot requires a SAW of sufficient amplitude to overcome the sloping potential and lift the electron into the channel. If the SAW amplitude is too large, it will carry the electron over the far barrier and out of the second dot. Thus there is a practical limit to the SAW amplitude for a given barrier/plunger combination, and for small-amplitude SAWs the dot needs to be raised towards the channel potential. Figure~\ref{FIG5}b shows how depopulation of LQD changes with SAW power and plunger voltage ($V_{\textrm{LP}}$). The potential slope from LQD up to the channel decreases as $V_{\textrm{LP}}$ increases, allowing smaller-amplitude SAWs with a shallower gradient to lift electrons from the dot.
Thus the onset of depopulation occurs along a diagonal line, between dashed lines in Fig.~\ref{FIG5}b and depopulation of the deeper dots requires larger amplitude SAWs.

 The pulse width of the SAWs may be varied instead of the power. It has previously been shown\cite{Kataoka2007} that a SAW can be used to modulate the barriers to an isolated dot causing population and depopulation of the dot in a probabilistic process that required many cycles to ensure $>50$\% probability of depopulation. Figure~\ref{FIG5}c shows how SAW pulse width, i.e. the number of attempts or SAW minima, affects depopulation of LQD.

Applied pulses are not reproduced exactly in the SAW pulses due to bandwidth limitations of the transducers,
pulses longer than $14$\,ns should vary only in duration and not in peak amplitude.
At $10.0$\,ns ($27.7$ cycles) the reduction in pulse amplitude due to transducer bandwidth is visible at the lower plunger voltages where electrons cannot be depopulated. By $12.6$\,ns ($34.9$ cycles), just $\sim7$ cycles more, depopulation is seen across almost the full range and, as expected, at $14.5$\,ns the SAW is able to remove electrons over the same range as pulses of much longer widths. From the rapid turn-on as the pulse width increases, with depopulation going from approximately zero to complete in just $12.5$ cycles, we can say that once a sufficient SAW amplitude is reached, depopulation occurs during the first few ($\sim7$) cycles of the pulse. Pulses applied to a transducer with a wider bandwidth (fewer fingers) would have shorter rise times, allowing this to be probed further.

 \begin{figure}[tbp]
\begin{center}
\includegraphics[width=8.5cm]{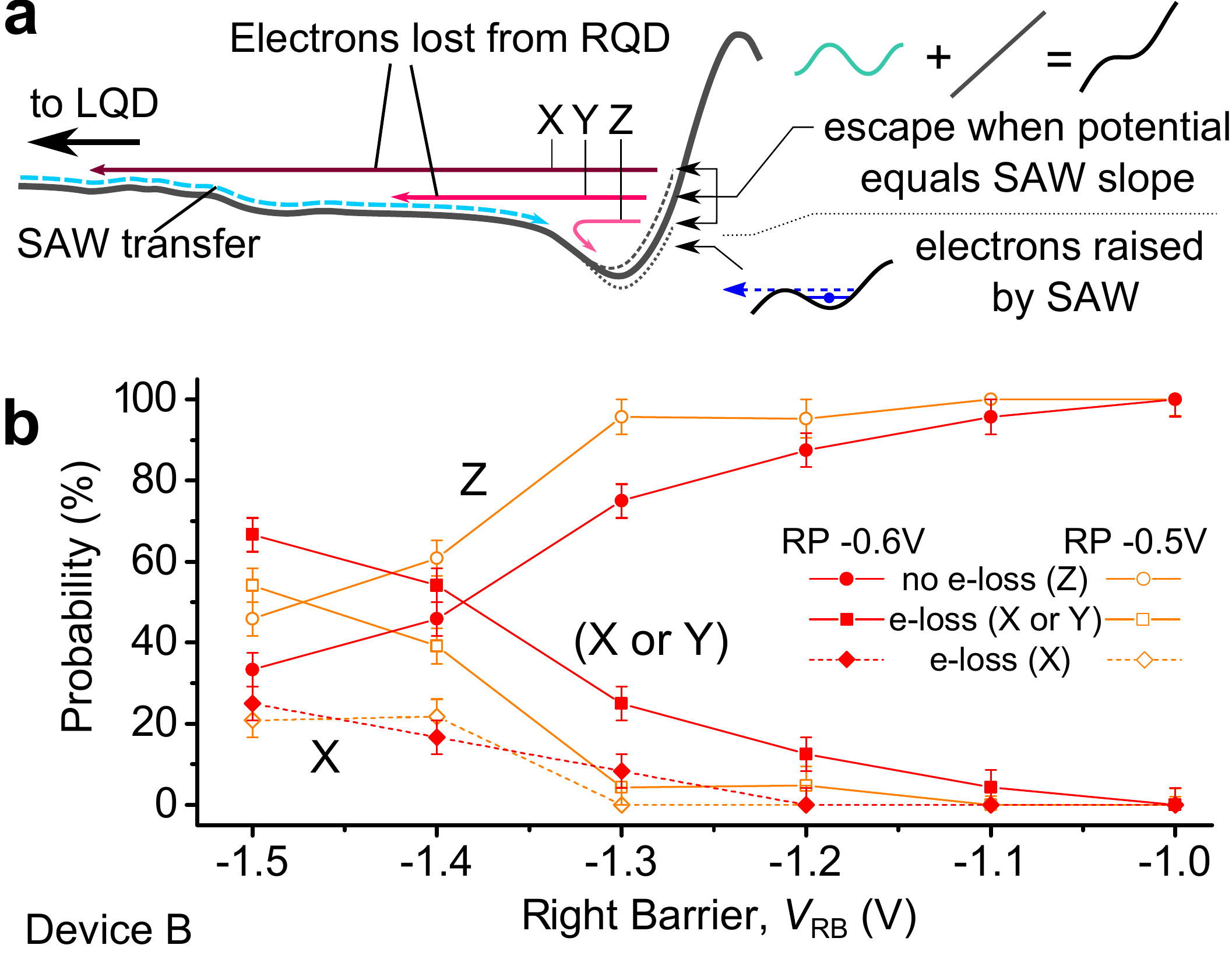}\\
\caption{\textbf{Backscattering of electrons in RQD due to SAW from the left.}
  \textbf{a}, Electron in RQD will be lifted up the right barrier by SAW(L) until it either leaves the system or the underlying potential becomes too steep for SAW minimum to retain the electron. Electrons either remain in dot (Z), escape to traps in channel (Y) or escape to LQD (X).
  \textbf{b}, Probability of events X, (X or Y), and Z \textit{vs} barrier voltage. (Open symbols are for slightly deeper dot potential.) Threshold is evident at $\sim -1.3$\,V: $V_{RB} < -1.3$\,V escape to LQD is possible, $V_{RB} > -1.3$\,V escape to LQD prevented by channel potential. Error bars show one standard deviation.}\label{FIG6}
\end{center}
\end{figure}
This system also provides a method of investigating energy-loss mechanisms for electrons above the Fermi energy.
As a SAW minimum transfers an electron, it lifts it over bumps in the potential, raising and lowering its potential energy as necessary. However, when the potential slope exceeds the maximum SAW gradient, confinement is lost and a ``hot'' electron escapes backwards towards the channel (Fig.~\ref{FIG6}a). The energy at which this occurs depends on the underlying potential. Figure~\ref{FIG6}b shows how varying the right barrier voltage ($V_{\textrm{RB}}$) affects the escape probability and the initial energy of the escaping electron. An electron starts in RQD and a long ($300$\,ns) SAW pulse is sent from the \textit{left}. Electrons escaping the SAW potential at a low energy will remain in the dot (Z), at higher energies they will escape to the channel (Y) and at energies above the channel maximum, they will reach LQD (X). During a pulse an X or Y electron may be returned to RQD and ``recycled'', with its ultimate position (LQD/trap in channel/RQD) being determined during the last part of the SAW pulse as the amplitude drops.
For $V_{\textrm{RB}} > -1.2$\,V, transfer to the channel is unlikely, with no electrons being transferred to LQD and the probability of staying in RQD (Z) $ > 90$\%. For $V_{\textrm{RB}} < -1.3$\,V the probability of leaving RQD (X or Y) increases to $>50$\% and the probability of escaping to LQD (X) reaches $25$\%. The open symbols are for a lower plunger voltage and show a reduced probability of transfer from RQD since RQD is now deeper.

Electrons with a large excess energy rapidly lose energy by emitting an optical phonon (of energy $36$\,meV) in about $1$\,ps \cite{Taubert2011}, comparable to the time for the electron to cross one QD. Below $36$\,meV electrons can only emit acoustic phonons with typical energies $\leq 0.1$\,meV and they also emit these phonons more slowly. In the low-energy limit this is on a $100$\,ns timescale.\cite{Fujisawa1998} The addition of a gate across the centre of the channel, capable of being pulsed at high frequencies,  would provide a method to investigate the emission of acoustic phonons by high-energy electrons.

This source of high-energy electrons may be of use in p-n junction devices as a way to controllably introduce single electrons into a region of holes as a single-photon source,\cite{Gell2006} without requiring negatively charged gates in close proximity to the holes.

To be useful in a quantum information circuit the transfer of an electron must not decohere its spin state.
Coherent transfer of a collection of spins has been demonstrated over $70$\,$\mu$m (for a particular wafer orientation) with the potential to extend this much further,\cite{Stotz2005} and coherent oscillations of charge have been shown over a sub-micron distance.\cite{Kataoka2009} Fluctuations in the magnetic field created by nuclear spins ($B_{\rm Nuc}$) are the main cause of dephasing in static QDs however in moving SAW quantum dots an electron samples many different local $B_{\rm Nuc}$ fields spending only a brief time in each. The average $B_{\rm Nuc}$, and so dephasing, is reduced by three orders of magnitude due to the motion of the SAW (more details of dephasing mechanisms are given in supplementary information). It is therefore likely that coherent transfer of spins is achievable and dephasing will actually be suppressed during the transfer.

In an ideal QD network, with a perfectly smooth potential, an electron could simply be allowed to ``roll'' from an elevated starting dot down to the second dot. In practice, the potential is far from perfect and irregularities from the background potential would make this method of transfer highly unreliable. A pulse of surface acoustic waves, however, can be used to temporarily modulate the channel, assisting the transfer in a peristalsis-like movement, the amplitude of which can be tuned to the minimum required to overcome desired obstacles, allowing on-demand removal and delivery of single electrons between distant quantum dots in a manner that should be compatible with many of the quantum computing proposals based on electron spin states in semiconductors.

\small
\textbf{Methods:}
The 2DEG is formed at the interface of a GaAs/AlGaAs heterostructure, before depletion, carrier density $1.6\times 10^{11}$\,cm$^{-2}$, mobility $1.8\times 10^{6}$\, cm$^{2}/Vs$. Device and transducers patterned by electron-beam lithography. All measurements done at $300$\,mK. RF signals are applied to the transducers with Agilent 8648D source (external modulation option). To prevent Bragg reflections transducers are of double-element design,\cite{Morgan2007} with $30$ pairs of fingers. Detector circuits share a common source with $\sim 1$\,mV DC bias. In device B RQD was adjusted between capture and transfer positions to aid depopulation by the weaker SAW(R). This adjustment shifts the dot minimum relative to the right detector, making the return steps smaller. Gate setup time between traces was $2-8$\,s. Applied RF power in Fig.~\ref{FIG1}d is 10dBm (SAW(L)) and 18dBm (SAW(R)), attenuation from source to transducers $10$\,dB (SAW(L)), $20-30$\,dB (SAW(R)).

\bibliographystyle{naturetitles}

\begin{thebibliography}{10}

\bibitem{Loss1998}
Loss, D. and DiVincenzo, D.~P.
\newblock Quantum computation with quantum dots.
\newblock {\em Phys. Rev. A}{ \bf 57}, 120 (1998).

\bibitem{Barnes2000}
Barnes, C. H.~W., Shilton, J.~M., and Robinson, A.~M.
\newblock Quantum computation using electrons trapped by surface acoustic
  waves.
\newblock {\em Phys. Rev. B}{ \bf 62}, 8410 (2000).

\bibitem{Petta2005}
Petta, J.~R., \textit{et al.}
\newblock Coherent manipulation of coupled electron spins in semiconductor
  quantum dots.
\newblock {\em Science}{ \bf 309}, 2180 (2005).

\bibitem{Pioro-Ladriere2008}
Pioro-Ladri\`{e}re, M., \textit{et al.}
\newblock {Electrically driven single-electron spin resonance in a slanting
  Zeeman field.}
\newblock {\em Nature Phys.}{ \bf 4}, 776 (2008).

\bibitem{Elzerman2004}
Elzerman, J.~M., \textit{et al.}
\newblock Single-shot read-out of an individual electron spin in a quantum dot.
\newblock {\em Nature}{ \bf 430}, 431 (2004).

\bibitem{Hanson2005}
Hanson, R., \textit{et al.}
\newblock Single-shot readout of electron spin states in a quantum dot using
  spin-dependent tunnel rates.
\newblock {\em Phys. Rev. Lett.}{ \bf 94}, 196802 (2005).

\bibitem{Morello2010}
Morello, A., \textit{et al.}
\newblock Single-shot readout of an electron spin in silicon.
\newblock {\em Nature}{ \bf 467}, 687 (2010).

\bibitem{Hanson2007}
Hanson, R., Kouwenhoven, L.~P., Petta, J.~R., Tarucha, S., and Vandersypen, L.
  M.~K.
\newblock Spins in few-electron quantum dots.
\newblock {\em Rev. Mod. Phys.}{ \bf 79}, 1217 (2007).

\bibitem{Field1993}
Field, M., \textit{et al.}
\newblock Measurements of Coulomb blockade with a noninvasive voltage probe.
\newblock {\em Phys. Rev. Lett.}{ \bf 70}, 1311 (1993).

\bibitem{Kataoka2006PRB}
Kataoka, M., Barnes, C. H.~W., Beere, H.~E., Ritchie, D.~A., and Pepper, M.
\newblock Experimental investigation of the surface acoustic wave electron
  capture mechanism.
\newblock {\em Phys. Rev. B}{ \bf 74}, 085302 (2006).

\bibitem{Rahman2006}
Rahman, S., Kataoka, M., Barnes, C. H.~W., and Langtangen, H.~P.
\newblock Numerical investigation of a piezoelectric surface acoustic wave
  interaction with a one-dimensional channel.
\newblock {\em Phys. Rev. B}{ \bf 74}, 035308 (2006).

\bibitem{Kataoka2007}
Kataoka, M., \textit{et al.}
\newblock Single-electron population and depopulation of an isolated quantum
  dot using a surface-acoustic-wave pulse.
\newblock {\em Phys. Rev. Lett.}{ \bf 98}, 046801 (2007).

\bibitem{Taubert2011}
Taubert, D., \textit{et al.}
\newblock Relaxation of hot electrons in a degenerate two-dimensional electron
  system: Transition to one-dimensional scattering.
\newblock {\em Pyhs. Rev. B}{ \bf 83}, 235404 (2011).

\bibitem{Fujisawa1998}
Fujisawa, T., \textit{et al.}
\newblock Spontaneous emission spectrum in double quantum dot devices.
\newblock {\em Science}{ \bf 282}, 932 (1998).

\bibitem{Gell2006}
Gell, J.~R., \textit{et al.}
\newblock Surface-acoustic-wave-driven luminescence from a lateral p-n
  junction.
\newblock {\em Appl. Phys. Lett.}{ \bf 89}, 243505 (2006).

\bibitem{Stotz2005}
Stotz, J. A.~H., Hey, R., Santos, P.~V., and Ploog, K.~H.
\newblock Coherent spin transport through dynamic quantum dots.
\newblock {\em Nature Mater.}{ \bf 4}, 585 (2005).

\bibitem{Kataoka2009}
Kataoka, M., \textit{et al.}
\newblock Coherent time evolution of a single-electron wave function.
\newblock {\em Phys. Rev. Lett.}{ \bf 102}, 156801 (2009).

\bibitem{Morgan2007}
Morgan, D.
\newblock {\em Surface Acoustic Wave Design.}
\newblock (Academic Press, London, 2007).

\end{thebibliography}


\end{document}


\title{Supplementary Information:\\On-demand single-electron transfer between distant quantum dots}
\author{R.~P.~G.~McNeil}
\author{M.~Kataoka\footnote{Current address: National Physical Laboratory, Hampton Rd, Teddington, Middlesex, TW11 0LW, United Kingdom.}}
\author{C.~J.~B.~Ford}
\author{C.~H.~W.~Barnes}
\author{D.~Anderson}
\author{G.~A.~C.~Jones}
\author{I.~Farrer}
\author{D.~A.~Ritchie}
\affiliation{Cavendish Laboratory, University of Cambridge, JJ Thomson Avenue, Cambridge CB3 0HE, United Kingdom }
\maketitle

\begin{figure*}[th]
  \centering
  \includegraphics[width=15cm]{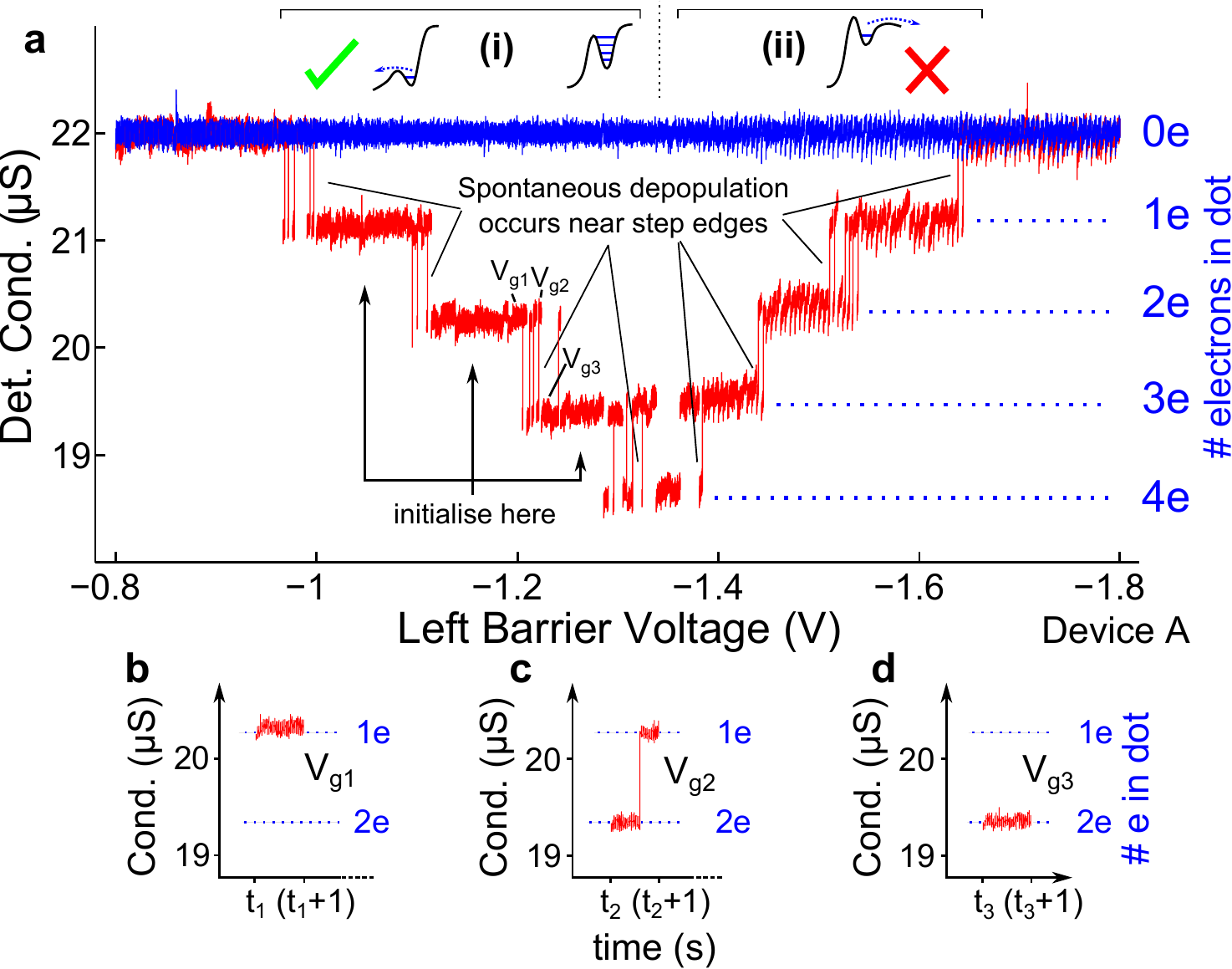}\\
  \label{Fig:S1}
\end{figure*}

{\noindent Fig~S1: \textbf{Setting up the QD and detector.} (\textbf{a}), Detector conductance \textit{vs} final barrier voltage. Data consists of many individual traces for $1$\,s intervals following initialisation as in Fig.~1b(i-iv \& v-viii). The individual traces are plotted aligned to the final left-barrier voltage. Vertical steps in conductance mark spontaneous depopulation of the dot. (\textbf{b-d}), Three examples of 1\,s long traces at voltages the ($V_{\rm g1}$, $V_{\rm g2}$ \& $V_{\rm g3}$) shown in (a).  Negative charge in the region of the detector channel suppresses conductance and so the stepwise separation between the blue (\textit{empty}) and red (\textit{occupied}) traces shows that the dot may be initialised with 0, 1, 2, 3 or 4 electrons as labelled in (a). (\textbf{i \& ii}) Diagrams of left dot potential at small and large barrier voltages.  At low barrier voltages (i) extra electrons depart to the left reservoir as intended, but at larger barrier voltages (ii) extra electrons escape to the channel, which is undesirable. Therefore the dots are  initialised on the left side of this inverted step pyramid.
The time taken to initialise one or more electrons in the dot was determined by the low-pass filters used on the gates (time constant $= 0.1$\,s), In later experiments this was reduced to 10$\mu$s and further reduction is possible.}\\

{\bf Setting up QD and detector:} Figure S1 illustrates the initialisation and characterisation of a quantum dot and its associated quantum point contact detector.\\

\textbf{Adjusting detector with gate voltage:} The detectors are set so that the detector conductance is on the steepest part of riser to the first 1D conduction plateau ($\sim 20-30\,\mu$S in our devices). As the detectors are sensitive to both neighbouring charge on the dot and the gates, they must be adjusted when changing other gates to keep them in the sensitive region. Calibration of occupancy is done with plots like the step pyramid above (Fig.~S1) and for a given gate tuning the expected step height produced by the arrival/departure of one or more electrons can be read from the step-pyramid plot.
In Fig.~S1 the detector conductance was tuned before each pair of traces but in general when plotting behaviour \textit{vs} a range of final gate voltages, the detector voltages were adjusted by linear interpolation between \textit{empty} calibration values at low and high gate voltages. In the transfer experiments the detector voltages have preset values which are fixed for the duration of the experiment.\\

\textbf{Ensuring the channel is empty:}
Depletion of the channel is confirmed at the start of a set of measurements by sending a series of SAW pulses from the left or right transducer to sweep the channel. Any remaining electrons would appear in either RQD or LQD.\\

\textbf{Power dependence:} (Device B)
At high and low powers Device B exhibited similar power-dependent behaviour to that of Fig.~3a, but at intermediate powers and plunger voltages electron transfer did not occur, probably due to an impurity in the static dot making the background potential steeper. However, the fidelity of the transfer process was \emph{not} affected, with electrons either staying in the first dot or being transferred to the second.\\

\textbf{Transducer bandwidth:}
The transducer bandwidth is the limiting bandwidth in the system. The geometry of the transducers sets a minimum rise-time of $\sim 11.6$\,ns during which the amplitude increases linearly, on top of the $2.1$\,ns rise time of the RF source, and a comparison of measured transducer bandwidth with a reflection-free $\delta$-function model\cite{Morgan2007} suggests that residual reflections under the transducers increase the rise time by less than $4$\%.\\

\textbf{Suppressed dephasing of SAW-driven dot due to spin-orbit coupling and nuclear spins:}
Spin-orbit coupling occurs when an electron moves in an electric field. The random electric field due to the impurity potential causes random spin rotation as described by the D'yakanov-Perel' (DP) and Elliot-Yafet (EY) mechanisms but these are suppressed by the confinement in a quantum dot.\cite{Hanson2007} In a SAW-driven quantum dot similar confinement is provided by the SAW and channel potentials and so this suppression also occurs.

However, because the dot is moving (at the speed of sound), there will be additional contributions to the spin rotation.
\begin{itemize}

\item{} Bulk inversion asymmetry in the GaAs (Dresselhaus term) and structural inversion asymmetry due to the electric field at the heterointerface (Rashba term) cause a fixed rotation that can be predicted and allowed for.

\item{} The electric field due to the impurity potential,  which appears as a magnetic field to the moving charge. This random field is 50 times smaller than the electric field at the heterointerface and so its contribution to dephasing should be insignificant. Additionally the effect of this ``impurity field'' is reduced further by the fact that it is approximately symmetric with the potential sloping both up and down for each impurity leading to significant cancellation.
    Although it will not be predictable in advance this rotation will be fixed and so subsequent calibration will be possible.

\item{}
EY and DP apply in 2D as the electron makes a random walk through the system. The SAW and gate potentials in our device mean the path of an electron is not random, as demonstrated by reference~\cite{Kataoka2009}.
A small random variation in the velocity occurs due to fluctuations in the position of the dot minimum. However, each electron transported through the channel will experience the same scattering potential and so this too should give a fixed rotation. If the device were made in an induced heterostructure then the background due to doping would be reduced.

\item{} In gallium arsenide the ensemble of $N$ nuclear spins encompassed by the electron wavefunction creates an effective magnetic field $B_{\rm Nuc}$ (Overhauser field) in which the electron spin precesses, through the hyperfine interaction. If all nuclear spins were aligned, the field would be of order 5\,T, but random alignment means the field is reduced by a factor $\sqrt{N}$, where $N$ is of order $10^6$ to $10^7$ for a dot diameter of 30--100\,nm and height 20--30\,nm. Fluctuations in $B_{\rm Nuc}$ cause uncertainty in the precession rate of the electron giving dephasing on a timescale $\tau_{\rm prec}$.
In static (0D) dots much work has been done on reducing dephasing due to fluctuations in the nuclear bath. In 2D systems the hyperfine interaction has a much reduced effect as itinerant electron wavefunctions are spread over a larger number of nuclei. In a static dot the mean field is determined by O($10^6$) nuclei whereas the field for a moving electron is that of all the nuclei it passes. Since fluctuations in the Overhauser field go as $\sqrt{N}$ the effective field is reduced by this \emph{motional narrowing}\cite{Slichter1989}, and so the very act of transporting the electron will reduce the Overhauser-induced spin precession (and so dephasing).
The dephasing time $T_2$ in a SAW-driven quantum dot can be shown to increase by a factor ($\tau_{\rm prec} \times v_{\rm SAW}/d$) where $v_{\rm SAW}$ is the SAW velocity and $d$ is the dot diameter. In our devices this factor would be 2000--7500, taking $\tau_{\rm prec}=83$\,ns\cite{Hanson2007}. During transfer, dephasing by interaction with $B_{\textrm{Nuc}}$ will therefore be heavily suppressed probably to the extent that other dephasing mechanisms will be the limiting factor.
\end{itemize}

\textbf{Adiabaticity of transport and capture by dynamic dot:}
An electron trapped in a SAW-driven quantum dot feels the SAW potential at all times, and energy moves between the SAW and the electron as the electron moves over a potential barrier---the SAW does work on the electron to increase its potential energy as it rises up the slope and the electron does work on the SAW as it moves down the other side. If the dot confinement is unchanged this process is adiabatic, i.e. the electron will stay in its ground state.
Raising and lowering a dot such that the minimum of the potential does not change its position does not leave the electron in an excited state provided that the rate of change in the shape of the confinement potential is slow. This type of manipulation is done in many double-dot and single-dot electron-spin-resonance (ESR) experiments.\cite{Hanson2007} In a similar manner, a SAW-driven electron will move adiabatically along the channel whilst experiencing a fluctuating impurity potential provided the variation in confinement is slow.
An electron in a SAW minimum that approaches a slope will slow briefly (experiencing a small change in momentum) at the base of the slope and will speed up briefly at the top. This can leave the electron in a partially excited state, and was the cause of the coherent charge oscillations seen in ref~\cite{Kataoka2009}, but for a sufficiently smooth transition it should be minimal.\\ 

\textbf{A note on transducer screening:}
Pulsing the transducers separates the electromagnetic (EM) wave radiated by the transducers from the mechanical wave that generates the piezoelectric potential modulation.
  A short microwave pulse of a few GHz would be (for a suitable Zeeman spin-splitting) like an off-resonance electron-spin-resonance (ESR) pulse and there should be no net spin rotation. Since the mechanical SAW wave and the EM wave are separated in time it may be possible to use the EM wave as a source of microwave photons to drive ESR transitions in specific static dots (tuned with local magnetic fields\cite{Pioro-Ladriere2008,McNeil2010}---resonant $B$ field $\sim 0.5$\,T) before the mechanical SAW arrives. However, this is not yet desirable and so our devices are screened by being in a cut-off waveguide.
The free space EM wavelength is $\sim10$\,cm and the device geometry allows it to be placed in a waveguide $\sim 1-2$\,mm across (cutoff frequency $\sim 60$\,GHz) so the free-space waves are heavily attenuated. The large transducer--device separation (1\,mm) also means that screening of the transducers can be achieved with a closely positioned superconducting or normal metal screen above the transducers or device (or both) to limit the microwave power reaching the QDs.

\bibliographystyle{naturetitles}